\documentclass[aps,pra,twocolumn,10pt,superscriptaddress,nofootinbib]{revtex4-1} 

\usepackage{graphicx}
\usepackage{amssymb}
\usepackage{amsmath}
\usepackage{verbatim}
\usepackage[colorlinks=true,citecolor=blue,linkcolor=blue]{hyperref}
\usepackage{dsfont}
\usepackage{subfigure}
\usepackage{natbib}

\newcommand{\trace}[1]{\mathrm{Tr}\left[#1\right]}
\newcommand{\tracesub}[2]{\mathrm{Tr}_{#2}\left[#1\right]}

\newcommand{\Udes}{U_{\mathrm{F}}}
\newcommand{\ns}{\mathrm{ns}}
\newcommand{\bra}[1]{\left<#1\right|}
\newcommand{\ket}[1]{\left|#1\right>}

\begin{document}

\title{Single qubit gates in frequency-crowded transmon systems}

\author{R. Schutjens}
\affiliation{Quantum Transport, Delft University of Technology, 2628 CJ Delft, The Netherlands }
\affiliation{Theoretical Physics, Universit\"{a}t des Saarlandes, 66123 Saarbr\"{u}cken, Germany}
\author{F. Abu Dagga}
\author{D. J. Egger}
\affiliation{Theoretical Physics, Universit\"{a}t des Saarlandes, 66123 Saarbr\"{u}cken, Germany}
\author{F. K. Wilhelm}
\affiliation{Theoretical Physics, Universit\"{a}t des Saarlandes, 66123 Saarbr\"{u}cken, Germany}
\affiliation{IQC and Department of Physics and Astronomy, University of Waterloo, Ontario N2L 3G1, Canada}

\date{\today}

\begin{abstract}
Recent experimental work on superconducting transmon qubits in 3D cavities show that their coherence times are increased by an order of magnitude compared to their 2D cavity counterparts. However to take advantage of these coherence times while scaling up the number of qubits it is advantageous to address individual qubits which are all coupled to the same 3D cavity fields. 
The challenge in controlling this system comes from spectral crowding, where leakage transition of qubits are close to computational transitions in other. Here it is shown that fast pulses are possible which address single qubits using two quadrature control of the pulse envelope while the DRAG method of Refs. \cite{Motzoi_PRL_103_110501, Gambetta_PRA_83_012308} alone only gives marginal improvements over the conventional Gaussian pulse shape. On the other hand, a first order result using the Magnus expansion gives a fast analytical pulse shape which gives a high fidelity gate for a specific gate time, up to a phase factor on the second qubit. Further numerical analysis corroborates these results and  yields to even faster gates, showing that leakage state anharmonicity does not provide a fundamental quantum speed limit. 
\end{abstract}

\maketitle

\section{Introduction}
Superconducting qubits are a promising candidate for the realization of a quantum computer \cite{Makhlin01,Insight,You05b,Reed_NatPhys_482_382,Lucero_NatPhys_8_719}, owing in large parts to the success of circuit QED (CQED), where those qubits are coupled to microwave resonators \cite{Blais_PRA_69_062320,Schoelkopf08,You11}. There is a multitude of designs of such qubits \cite{Insight}. 

A key challenge for implementing quantum computing in the solid state is decoherence from uncontrolled degrees of freedom. Decoherence sources range from the electromagnetic environment \cite{Koch_PRA_76_042319} to sources inherent to the material \cite{PhysRevLett.95.210503}. Remarkably, many of the material sources could be mitigated by changes in the circuit layout such as the optimum working point first embodied in the Quantronium \cite{Vion02,Makhlin04,Rebentrost09B,PhysRevB.76.174516} and later in the Transmon \cite{Koch_PRA_76_042319} and the 
3D-Transmon \cite{Paik_PRL_107_240501,Rigetti_PRB_86_100506}.
Coherence times have been improved by going from the two dimensional implementation of a qubit interacting with a stripline resonator \cite{Blais_PRA_69_062320} to a three dimensional system \cite{Paik_PRL_107_240501,Rigetti_PRB_86_100506}. In the latter, a single Josephson junction transmon qubit \cite{Koch_PRA_76_042319,Schuster06} is placed inside a 3D cavity and addressed with the surrounding microwave field. What is common to these approaches is the trade-off of coherence against control flexibility and ultimately operation speed. While this has been studied in single Quantronium \cite{PhysRevB.76.174516} the precise trade-off is not fully understood in samples containing multiple qubits let alone multiple 3D transmons.

The gain in coherence times comes at a cost in controllability. This is strongly felt when more than one qubit is in the cavity. To create single qubit operations each qubit must be addressed individually requiring them to have significantly different energy splitting between the ground and first excited state. Spectral crowding refers to transitions coming too close to address them individually. Now with the limited control, even if the logical transitions are well-spaced, crowding can occur between logical and leakage transition, e.g., if the logical transition of first qubit is close in frequency to the leakage transition, the transition between a computational and a non-computational state,  of the second qubit. Thus when performing, e.g., an $\hat{X}$ gate on first qubit leakage to second qubit's $|2\rangle$ state will occure. Although high fidelity gates have been demonstrated with single junction transmsons in the 2D architecture \cite{Chow_PRL_109_060501} spectral crowding will limit the gate fidelity in 3D architectures. 
In order to mitigate spectral overlap, the Derivative Removal by Adiabatic Gate (DRAG) 
technique has been developed \cite{Motzoi_PRL_103_110501,Gambetta_PRA_83_012308}. 
We will apply this technique to the problem at hand and show that on its own it is of limited success. We will then combine DRAG with sideband drive to show a possibility to do these single-qubit gates fast. 

In this work we thus address the issue of spectral crowding with optimal control theory methods.  To better illustrate the problem and show the effectiveness of the analytical pulses we introduce specific gate fidelity functions in section \ref{sec:gates}. In section \ref{sec:drag} we demonstrate the limitations of the DRAG technique alone for this problem. We then present an analytical pulse, found through the Magnus expansion \cite{Warren84}, capable of minimizing leakage out of the computational subspace of both qubits in section \ref{sec:Magnus}. We then, in section \ref{sec:num}, show pulses obtained numerically that show similar characteristic but, with additional ingredients, improved fidelities. 

\section{System}
\label{sec:system}

Optimized superconducting qubits such as 3D transmons are well described by weakly anharmonic oscillators \cite{PhysRevLett.106.030502,Motzoi_PRL_103_110501}. A realistic model of the qubit has to take at least one extra non-computational level (a {\em leakage level}) into account \cite{Chow09b,Steffen03,Khani09}. 
This is reflected in the following Hamiltonian for two superconducting transmon qubits in a common 3D cavity 

\begin{equation}
\label{eq:H1}
\begin{split}
\hat H \left( t \right) &= \hat H_0+\hat H_C \left( t \right) \\
&= \sum_{k=1}^2 \left[ \omega_{k}\hat n_k+\Delta_k\hat\Pi_2^{(k)} \right] \\
&+\Omega \left( t \right) \sum_{j=1}^2 \left[\lambda_j^{\left(1\right)} \hat\sigma_{j,j-1}^{x \left(1\right)}+\lambda_j^{\left(2\right)} \hat\sigma_{j,j-1}^{x \left(2\right)} \right].
\end{split}
\end{equation}

The $0\leftrightarrow1$ transition frequency and number operator of qubit $k$ are, respectively, $\omega_k$ and $\hat n_k=\sum_jj\ket{j}\bra{j}^{(k)}$. 
We call the transition from the excited state $\ket{1}$ to the extra state $\ket{2}$ the leakage transition. It is detuned from $\omega_k$ by the anharmonicity $\Delta_k$. In the reminder of this work we assume $\Delta_1=\Delta_2=\Delta$. The projectors on the energy levels of transmon $k$ are $\hat\Pi_k^{\left(k\right)}=\left|j\right>\left<j\right|^{\left(k\right)}$. The terms coupling adjacent energy levels of qubit $k$ are
\begin{equation} \notag
 \hat\sigma_{j,j-1}^{x \left(k\right)}=\left|j\right>\left<j-1\right|^{\left(k\right)}+\left|j-1\right>\left<j\right|^{\left(k\right)}
\end{equation}
and
\begin{equation}\notag
 \hat\sigma_{j,j-1}^{y \left(k\right)}=i\left|j\right>\left<j-1\right|^{\left(k\right)}-i\left|j-1\right>\left<j\right|^{\left(k\right)}.
\end{equation}
$\Omega(t)$ is the drive field and is applied simultaneously to both qubits. The strength at which $\Omega(t)$ drives the $1\leftrightarrow2$ transition relative to the $0\leftrightarrow1$ is given by $\lambda_j^{(k)}$. Table \ref{table:2} show the variables and numerical values used in simulations. \footnote{These values were suggested to describe an experiment by Leo DiCarlo} 

\begin{table}[htbp!]
\caption{System parameters as shown in equation (\ref{eq:H1}).} 
\centering 
\begin{tabular}{l @{\hspace{2em}} c @{\hspace{2em}} c @{\hspace{2em}} l} 
\hline \hline 
& Qubit 1 & Qubit 2 & \\ [0.5ex] 
\hline
$\omega_k/ 2\pi$ & $5.508$ & $5.5903\:$ & $\mathrm{GHz}$   \\ 
$\Delta/ 2\pi$ &  $-350$ & $-350$ & $\mathrm{MHz}$\\
$\lambda_1^{(k)}$ & $1$& $1$ & \\
$\lambda_2^{(k)}$ & $\sqrt{2}$&$\sqrt{2}$ & \\ [1ex] 
\hline \hline 
\end{tabular}
\label{table:2} 
\end{table}

\begin{figure}[htbp!] \centering
 \includegraphics[width=0.35\textwidth]{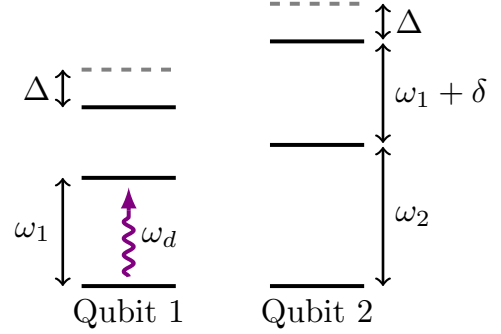}
 \caption{Level diagram of the two qubits. The driving field is set to have the same frequency as the $0\leftrightarrow1$ transition of first qubit which we wish to drive. Requiring that the same transition of the second qubit be far detuned results in its leakage transition being only slightly detuned by $\delta$ with $0\leftrightarrow1$ of first qubit. \label{Fig:Sketch}}
\end{figure}

Qubits are usually addressed by frequency selection through pulses tuned to the respective qubit level splitting. This is necessary whenever the control field cannot be selectively focused on individual qubits as is the case for multiple 3D transmons in the same cavity.
An eventual implementation of a quantum computer will consist of many such qubits, probably a whole register in one cavity.
The problem to distinguish different qubits can thus be seen as a problem of spectral crowding.
In transmon systems this can lead to the  $0\leftrightarrow1$
transition of the first qubit being very close to the $1\leftrightarrow2$ transition of the second qubit. The frequency difference of these two transitions is named $\delta$. With $\delta/2 \pi=45 \: \mathrm{MHz}$, 
the leakage transition of qubit two is closer to the driving fields frequency than the leakage transition of qubit one detuned by $\Delta/2 \pi=-350 \: \mathrm{MHz}$. The situation is depicted in Fig. \ref{Fig:Sketch}.

The second term in equation (\ref{eq:H1}) is the control Hamiltonian, described as a semiclassical dipolar interaction between the qubits and the classical cavity field
\begin{equation}
\Omega\left( t\right)=\Omega_X\left( t\right)\cos\left(\omega_d t\right)+\Omega_Y\left( t\right) \sin\left( \omega_d t\right).
\end{equation}
Both quadrature envelopes can be modulated separately. In the reminder of this work, we assume resonance between the drive and qubit 1, i.e. $\omega_d=\omega_1$.
Single quadrature pulses employ Gaussian shapes $\Omega_g$ due to their limited bandwidth \cite{Gambetta_PRA_83_012308}. To remove fast oscillating terms we move to another reference frame and invoke the rotating wave approximation (RWA). 
The transformation into an appropriate frame  is accomplished by the time-dependent unitary $\hat{R}$ that acts on the Hamiltonian as 
\begin{equation}
\hat H^R=\hat R\hat H\hat R^{\dagger}+i\dot{\hat R}\hat R^{\dagger}.
\label{eq:togglingframe}
\end{equation}
Here, $\hat R\left(t\right)=\left(\sum_j e^{-i \omega_j^{\left(1\right)}t}\hat \Pi_j^{\left(1\right)}\right)\otimes \left(\sum_j e^{-i \omega_j^{\left(2\right)}t}\hat\Pi_j^{\left(2\right)}\right)$. Transformations into this type of frame can lead to either the rotating frame with respect to the drive $\omega_d$ or the interaction frame by choosing $\omega_j^{\left(l\right)}=j \omega_d$, $\omega_j^{\left(l\right)}=j \omega^{\left(l\right)} +\Delta_j^{(l)}$ respectively. Here, we choose the former. In the rotating frame, we use the RWA to neglect the fast oscillating terms such as $\pm 2 \omega_d$, the system's original Hamiltonian given by (\ref{eq:H1}), is
\begin{equation}
\label{eq:HR}
\begin{split}
\hat H^R&= \Delta \hat\Pi_2^{\left(1\right)}+\left(\delta- \Delta \right)\hat\Pi_1^{\left(2\right)} + \delta\hat\Pi_2^{\left(2\right)}\\
&+\frac{\Omega_X \left( t \right)}{2}\sum_{j=1}^2 \left[\lambda_j^{\left(1\right)} \hat\sigma_{j,j-1}^{x \left(1\right)}+\lambda_j^{\left(2\right)} \hat\sigma_{j,j-1}^{x \left(2\right)} \right]\\
&+\frac{\Omega_Y \left( t \right)}{2}\sum_{j=1}^2 \left[\lambda_j^{\left(1\right)} \hat\sigma_{j,j-1}^{y \left(1\right)}+\lambda_j^{\left(2\right)} \hat\sigma_{j,j-1}^{y \left(2\right)} \right].
\end{split}
\end{equation}

\section{Single qubit gates}
\label{sec:gates}

We aim at applying, up to a global phase $\phi$, a gate on the first qubit without affecting the second one
\begin{equation}
\label{eq:Udes}
\hat\Udes =e^{i\phi}\hat U^{\left(1\right)}\otimes\mathds{1}.
\end{equation}
Unless otherwise specified $\hat U^{(1)}$ is an $\hat{X}$-gate. A specific control pulse of duration $t_g$ results in a final gate given by $\hat U\left(t_g\right)$. The fidelity with which a control pulse meets the target gate is measured by
\begin{equation}
\label{eq:fid}
\Phi=\frac{1}{d^2}\left|\trace{\hat\Udes^{\dagger}\hat U\left(t_g\right)}\right|^2,
\end{equation}
where $d$ is the dimension of the Hilbert space of the system. The trace is taken over the computational subspace consisting of $\{\ket{00},\ket{01},\ket{10},\ket{11}\}$. This takes leakage into account since leaving this subspace diminishes the matrix elements of the projected unitary \cite{Gambetta_PRA_83_012308, Rebentrost09}.

We will also investigate single-qubit gates that shift the phase of the second qubit. Such gates can be made more efficiently and we later show how to correct the phase.
Such gates can be studies using the reduced fidelity functions
\begin{equation}
\label{eq:fidi}
\Phi_{\ket{*,i}}=\frac{1}{2^2}\left|\tracesub{\hat \Udes^{\dagger}\hat U\left(t_g\right)}{\left\{\ket{0,i},\ket{1,i}\right\}}\right|^2.
\end{equation}
The trace is taken over states where the second qubit is exclusively in the $\ket{0}$ or $\ket{1}$. A gate producing a good $\Phi_{\ket{*,i}}$ has qubit 2 starting and ending in state $\ket{i}$. The average of the $\Phi_{\ket{*,i}}$'s gives a fidelity function insensitive to the phase of the second qubit
\begin{equation}
\label{eq:fidavg}
\Phi_{\mathrm{avg}}=\frac{1}{2}\left(\Phi_{\ket{*,0}}+\Phi_{\ket{*,1}}\right).
\end{equation}
In other words, $\Phi_ {\rm avg}$  is maximal if $\hat{U}(t_g)$ (in the computational subspace of the two qubits) has the form
\begin{equation}
\label{eq:Udesfidavg}
\hat U(t_g)
=e^{i\alpha}
\begin{bmatrix}
  0& 1\\
  1 & 0
\end{bmatrix}
\otimes 
\begin{bmatrix}
  1 & 0\\
  0 & e^{i\left( \gamma-\alpha \right)}
\end{bmatrix}.
\end{equation}
For a given gate time the phase error can be calculated and subsequently corrected as this gate is not entangling. In fact, an entangling gate would be detected by deteriorating $\Phi_{\rm avg}$ and given that the qubit controls are local and the two qubits are uncoupled, no entanglement is generated. 

\section{Applying DRAG} \label{sec:drag}

The DRAG method \cite{Gambetta_PRA_83_012308,Motzoi_PRL_103_110501,Motzoi13} negates leakage to the $\ket{2}$ state with a two quadrature drive. Here we show that this method does not provide a sizable improvement over a single Gaussian envelope. We transform $\hat H^R$ a second time along the lines of eq. (\ref{eq:togglingframe}) using the transformation matrix 
\begin{equation}
\hat{V} \left(t\right)=\exp\left(-i\frac{\Omega_X}{2\beta}\sum_{j=1}^2\left[\lambda_j^{\left(1\right)} \hat{\sigma}_{j,j-1}^{y \left(1\right)}+\lambda_j^{\left(2\right)} \hat{\sigma}_{j,j-1}^{y \left(2\right)} \right]\right).
\end{equation}
This is the two-qubit version of the DRAG transformation \cite{Gambetta_PRA_83_012308,Motzoi13}. The parameter $\beta$ selects which transition is suppressed. A first order expansion in $\eta=\Omega_X\left(t\right)/\beta\ll1 $ gives
\begin{equation}
\label{eq:dragscheme}
\hat H^{V}=\hat H_{\mathrm{diag}}+\hat H_{\mathrm{Y}}+\hat H_{\mathrm{X}}^{(1)}+\hat H_{\mathrm{X}}^{(2)}
\end{equation}
The diagonal terms are of $O(\eta^2)$, hence $\hat H_\mathrm{diag}$ is neglected on our level of approximation. $\hat H_{\mathrm{Y}}$ contains a term generated by the time-derivative in eq. (\ref{eq:togglingframe}) as well as the $Y$ drive
\begin{equation}
\hat H_{\mathrm{Y}}=\left(\frac{\Omega_Y\left(t\right)}{2}+\frac{\dot{\Omega}_X\left(t\right)}{2 \beta} \right)\sum_{j=1}^2\left[\lambda_j^{\left(1\right)} \hat\sigma_{j,j-1}^{y \left(1\right)}+\lambda_j^{\left(2\right)} \hat\sigma_{j,j-1}^{y \left(2\right)} \right].
\end{equation}
$\hat H_\mathrm{Y}$ can be suppressed by choosing $\Omega_Y\left(t\right)=-\dot{\Omega}_X\left(t\right)/\beta$. This is the essence of the DRAG method \cite{Motzoi_PRL_103_110501}. The last two terms respectively drive the first and second qubit according to
\begin{align} \notag
\hat{H}_{\mathrm{X}}^{(1)}(t)=&~\Omega_X (t)\hat{\sigma}_{10}^{x \left(1\right)}+\lambda \frac{\beta-\Delta}{2\beta}\Omega_X(t) \hat{\sigma}_{21}^{x \left(1\right)} \\ &+\frac{\lambda \Delta}{8\beta^2}\Omega_X(t)^2\hat{\sigma}_{20}^{x(1)}, \notag \\
\hat{H}_{\mathrm{X}}^{(2)}(t)=&~\eta \frac{\beta-\delta+\Delta}{2\beta}\Omega_X(t) \hat{\sigma}_{10}^{x \left(2\right)}+\eta \lambda \frac{\beta-\delta}{2\beta}\Omega_X(t) \hat{\sigma}_{21}^{x \left(2\right)} \notag \\
&+\frac{\eta^2\lambda\Delta}{8\beta^2}\Omega_X^2(t)\hat{\sigma}_{20}^{x\left(2\right)}. \notag
\end{align}

\begin{figure}[htbp!] \centering
\includegraphics[width=1\columnwidth]{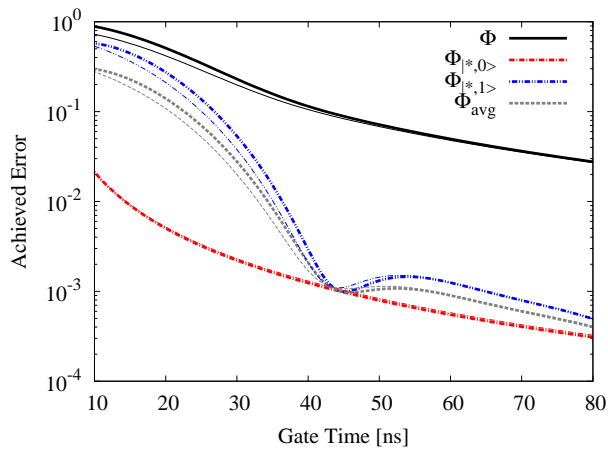}
\caption{Error for a single control with a Gaussian pulse shape as a function of gate time and a single quadrature (thin lines) and for the DRAG method with $\beta=\Delta$ (thick lines). The DRAG method gives only marginal improvements over the single quadrature Gaussian pulse shape for $\Phi_{\mathrm{avg}}$ which is slightly lower at the dip around 42 ns. The DRAG solution shown here is the optimal from picking $\beta \: \mathrm{\epsilon} \: \{\Delta,\delta,\delta-\Delta\}$.}
\label{fig:gaussanddrag}
\end{figure}

Depending on the value of $\beta$ a specific off resonant transition can be suppressed. If $\beta=\delta$ the second qubit leakage transition is removed. However, since $\delta<\Delta$ (by a factor $>7$ for the numbers in table \ref{table:2}) the compensation field $\Omega_Y$ becomes large and
strongly drives the other leakage transitions, i.e., introduces errors of a size comparable to what it is suppressing. Note, that for fast pulses with $\beta=\delta$ the perturbation expansion in \cite{Gambetta_PRA_83_012308,Motzoi_PRL_103_110501,Motzoi13}. naturally breaks down. Selecting $\beta=\Delta$ suppresses the leakage transition of the first qubit, but does not solve the leading spectral crowding issue based on the smallness of $\delta$. We are explicitly highlighting this in 
figure \ref{fig:gaussanddrag}. It shows the fidelity, as a function of gate time, for the single quadrature Gaussian (thin lines) and DRAG (thick lines) solutions with $\beta=\Delta$. 

The difference between the fidelity function $\Phi$, eq. (\ref{eq:fid}), and the special fidelity functions $\Phi_{\ket{*,i}}$ and $\Phi_{\mathrm{avg}}$, eq. (\ref{eq:fidi}) and (\ref{eq:fidavg}), show that while it is difficult to perform an $X$ gate on qubit 1 without affecting qubit 2, we can implement a high fidelity $\hat{X}$ gate with an additional phase shift on the other qubit for $t_g>42\ns$. This marks a time limitation that for DRAG alone to produce a high-fidelity gate the time needs to be at least on the boundaries of the adiabatic regime.

\section{Magnus expansion}
\label{sec:Magnus}

Here we show how to find an improved pulse capable of performing the desired gate faster and with better fidelity. The full effect of system and Hamiltonian is described by the time evolution operator
\begin{equation}
\label{eq:TOU}
\hat U\left(t_g\right)=\mathds{T} \exp\left\{-i \int\limits_0^{t_g} \mathrm{d}t\, \hat H(t)\right\}
\end{equation}
where $\mathds{T}$ is the time-ordering operator.
This can in general not be computed in closed form even for driven two-state systems with notable exceptions \cite{Gangopadhyay10}. Still being unitary, the solution of equation (\ref{eq:TOU}) can be written as the exponential of an Hermitian matrix \cite{Warren84}. An expansion in this effective Hamiltonian gives the Magnus expansion 
\begin{equation}
U\left(t_g\right)=e^{-i \sum_k \hat\Theta_k \left(t_g\right)}.
\label{eq:MagnusExpansion}
\end{equation}
The equation above still requires exponentiating a matrix. However the absence of time ordering considerably simplifies the derivation of an explicit expression for $\hat U$. The Magnus expansion is asymptotic. Here, it converges quickly as nested integrals lead to cancellations of fast oscillating terms. The constraints on the controls set by the zeroth order in the expansion will thus be most important. 

The first terms in the expansion are given by \cite{Warren84}
\begin{equation}
\begin{split}
\hat\Theta_0 \left(t_g\right)=& \int\limits_0^{t_g}\mathrm{d}t \hat H(t),\\
\hat\Theta_1 \left(t_g\right)=& -\frac{i}{2} \int\limits_0^{t_g}\mathrm{d}t_2 \int\limits_0^{t_2}\mathrm{d}t_1 \left[\hat H\left(t_2\right),\hat H\left(t_1\right)\right].
\end{split}
\end{equation}
Here $\left[\hat H\left(t_2\right),\hat H\left(t_1\right)\right]$ is the commutator of the Hamiltonian at different times. Higher order terms in the expansion can be worked out as nested commutators similar as those shown above. 

We start with the system in the interaction frame (the transformation is given in section \ref{sec:system})
\begin{equation}
\label{eq:HI}
\begin{split}
\hat H^I&= \frac{\Omega_C}{2}\sum_{j=1}^2 \left[\lambda_j^{\left(1\right)} e^{-i\delta_j^{\left(1\right)} t}\ket{j-1}\bra{j}^{\left(1\right)}\right.\\
&+\left.\lambda_j^{\left(2\right)} e^{-i\delta_j^{\left(2\right)} t} \ket{j-1}\bra{j}^{\left(2\right)} \right]+\mathrm{h.c.}
\end{split}
\end{equation}
Here we have combined $\Omega_C=\Omega_X+i\Omega_Y$ and set $\delta_1^{\left(1\right)}=0, \delta_2^{\left(1\right)}=\Delta, \delta_1^{\left(2\right)}=\delta-\Delta$, and $\delta_2^{\left(2\right)}=\delta$. In the interaction frame, the Hamiltonian is purely off-diagonal and the desired gate is changed by a phase on the $\ket{1}$ state of the second qubit. This phase is known since any unitary transformation $\hat V\left(t\right)$, transforms the time evolution following $\hat U^V\left(t_g\right)=\hat V\left(t_g\right)\hat U\left(t_g\right)\hat V^{\dagger}\left(0\right)$. In equation (\ref{eq:UfM}) $\Udes$ transforms in this way.
If the zeroth order term is to implement the gate, the control problem becomes
\begin{equation}
\label{eq:UfM}
\hat \Udes=e^{-i\hat \Theta_0}=e^{-i\int_0^{t_g}\mathrm{d}t\, \hat H^I\left(t\right)}.
\end{equation}
As an aside, this highlights why $\Theta_0/t_g$ is often called the average Hamiltonian and $\sum_k \hat{\Theta}_k(t_g)/t_g$ the effective Hamiltonian in NMR \cite{Warren84}.
This and the form $\hat H^I$ imposes restrictions on the control $\Omega_C$
\begin{eqnarray}
\frac{1}{2}\int_0^{t_g}\mathrm{d}t \: \Omega_C&=&\pi \label{eq:FC1}
\\
\frac{1}{2}\int_0^{t_g}\mathrm{d}t \: e^{-i\Delta t}\Omega_C&=&0 \label{eq:FC2}\\
\frac{1}{2}\int_0^{t_g}\mathrm{d}t \: e^{-i\delta t}\Omega_C&=&0 \label{eq:FC3}\\
\frac{1}{2}\int_0^{t_g}\mathrm{d}t \: e^{-i\left(\delta-\Delta\right) t}\Omega_C&=&0 \label{eq:FC4}
\end{eqnarray}
These constraints are the Fourier transforms of the control evaluated at the different detunings in the system as is familiar
from spectroscopy at weak drive  \cite{Vold68,Hoult79,Warren84,Freeman98}- but here derived under intermediate-to-strong drive conditions. 
They state that the control should contain no power at the off resonant frequencies. If $\Omega_C$ is palindromic the complex conjugated equations are also satisfied. If equations (\ref{eq:FC1}-\ref{eq:FC4}) are met, the final unitary evolution will be $e^{i\phi}\hat\sigma_x\otimes\mathds{1}$.

\begin{figure}[htbp!] \centering
\includegraphics[width=0.95\columnwidth]{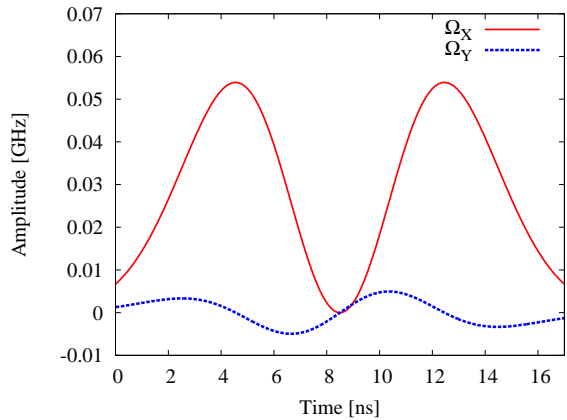}
\caption{Example of the control functions of equation (\ref{eq:wahwahcontrols}) for $t_g=17\ns$. The amplitude of $\Omega_x$ is somewhat smaller than for a Gaussian only pulse (which has been used in figure \ref{fig:gaussanddrag}). }
\label{fig:wahwahcontrols}
\end{figure}

So that the zeroth order implements the gate, higher order terms have to be zero. Here is an example of the first order term $\hat \Theta_1$. It only gives extra terms on the diagonal and the $0\leftrightarrow2$ transition. This calculation is quite involved and here is an example of the term involving $\ket{0,1}\bra{0,1}$ (neglecting terms oscillating faster than $\delta$)

\begin{equation}
\begin{split}
\langle 01|\hat\Theta_1\left(t_g\right)|01\rangle=\frac{1}{4}\int_0^{t_g}\mathrm{d}t_2 \int_0^{t_2}\mathrm{d}t_1 \Omega(t_1,t_2)\\ \left[1+\cos\left(\delta \left(t_1-t_2\right)\right)-\sin\left(\delta\left(t_1-t_2\right)\right)\right],
\end{split}
\end{equation}

with $\Omega(t_1,t_2)= \Omega_X\left(t_2\right)\Omega_Y\left(t_1\right) - \Omega_X\left(t_1\right)\Omega_Y\left(t_2\right)$. In the spirit of the Magnus expansion, all slow oscillating terms have the form above and are negligible if their integral is small. This suggest a control pulse where $\Omega_X$ is modulated with a sinusoidal function
\begin{equation}
\label{eq:wahwahcontrols}
\begin{split}
\Omega_X=&A_{\pi} e^{-\frac{1}{2\sigma^2}\left(t-\frac{t_g}{2}\right)^2}\left(1-A \cos\left[\omega_x\left(t-\frac{t_g}{2}\right)\right]\right), \\
\Omega_Y=&-\frac{1}{\beta}\dot{\Omega}_X.
\end{split}
\end{equation}
This is a Gaussian with added sideband modulation on the in-phase part $\Omega_x$ supplemented by DRAG on the quadrature $\Omega_y$. A frequency modulation with $\cos(\omega_x t)$ for a bandwidth of $\Omega_g<2 \omega_x$ can be seen as adding an effective drive at $\omega_x$ proportional to $\Omega_g$. This added drive can be used to counteract the population transfer of a specific transition. The absolute errors of eqs. (\ref{eq:FC2}-\ref{eq:FC4}) are minimized by varying $A,\: \omega_x,\: \beta$ yielding a pulse with a sideband modulation of $\delta/2$
\begin{equation}
\label{eq:wahwah}
\begin{split}
\Omega_X=&A_\pi e^{-\frac{18}{t_g^2}\left(t-\frac{t_g}{2}\right)^2}\left(1- \cos\left(\frac{\delta}{2}\left(t-\frac{t_g}{2}\right)\right)\right),\\
\Omega_Y=&-\frac{1}{2\Delta}\dot{\Omega}_X.
\end{split}
\end{equation}
Here we chose $\sigma=t_g/6$. The factor of 2 in the denominator of $\Omega_Y$ comes from the absence of control over the qubit frequency \cite{Gambetta_PRA_83_012308}. This is shown experimentally in ref. \cite{Chow10, Lucero10}. The pulse is shown in figure \ref{fig:wahwahcontrols} for $t_g=17\ns$ and other parameters given by the values in table \ref{table:2}. In order for the pulse to produce the $X$ gate $A_\pi$ should be chosen so that relation (\ref{eq:FC1}) is satisfied.

\subsection{Sideband modulation}
\label{sec:SBMdisc}
\begin{figure}[htbp!] \centering
\includegraphics[width=0.95\columnwidth]{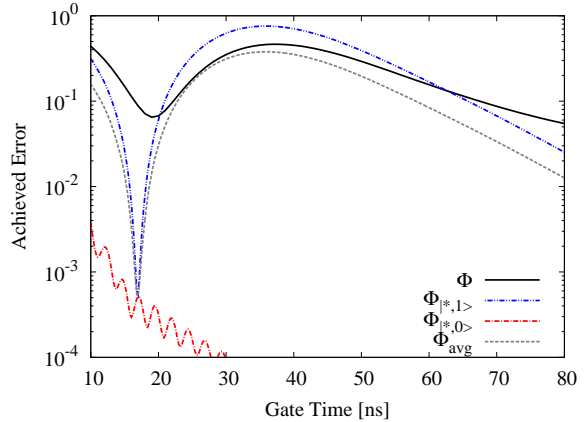}
\caption{Error as a function of gate time for the pulse with sideband modulation. The target gate is $\hat\sigma_x\otimes\mathds{1}$. At $t_g\sim17\ns$, $\Phi_{\mathrm{avg}}$ reaches a maximum. The gate fidelity functions are defined in equations (\ref{eq:fid}), (\ref{eq:fidi}) and (\ref{eq:fidavg}) respectively.}
\label{fig:wahwahvstg}
\end{figure}

The black line in figure \ref{fig:wahwahvstg} shows the error of pulse (\ref{eq:wahwah}) as function of gate time. Compared to the Gaussian and DRAG results, the error has a minimum ($~4\%$) at a shorter gate time, around $20\ns$. The reduced fidelity functions $\Phi_{\ket{*,i}}$ (red and blue lines) and $\Phi_{\mathrm{avg}}$ (gray line) give additional insight by allowing a phase shift on qubit 2. Comparing to figure \ref{fig:gaussanddrag}, it is seen that the sideband modulated pulse attains a high fidelity ($>99.9\%$) in less than half the time ($17~\mathrm{ns}$ compared to $42~\mathrm{ns}$) of the Gaussian or DRAG solutions. The $1 \leftrightarrow 2$ transition of the second qubit is still the limiting factor since the reduced error $1-\Phi_{\ket{*,1}}$ is always the biggest.  Nonetheless for a specific gate time a high fidelity is possible.

\begin{figure}[htbp!] \centering
  \subfigure{\label{Fig:subfiga}\includegraphics[width=0.95\columnwidth]{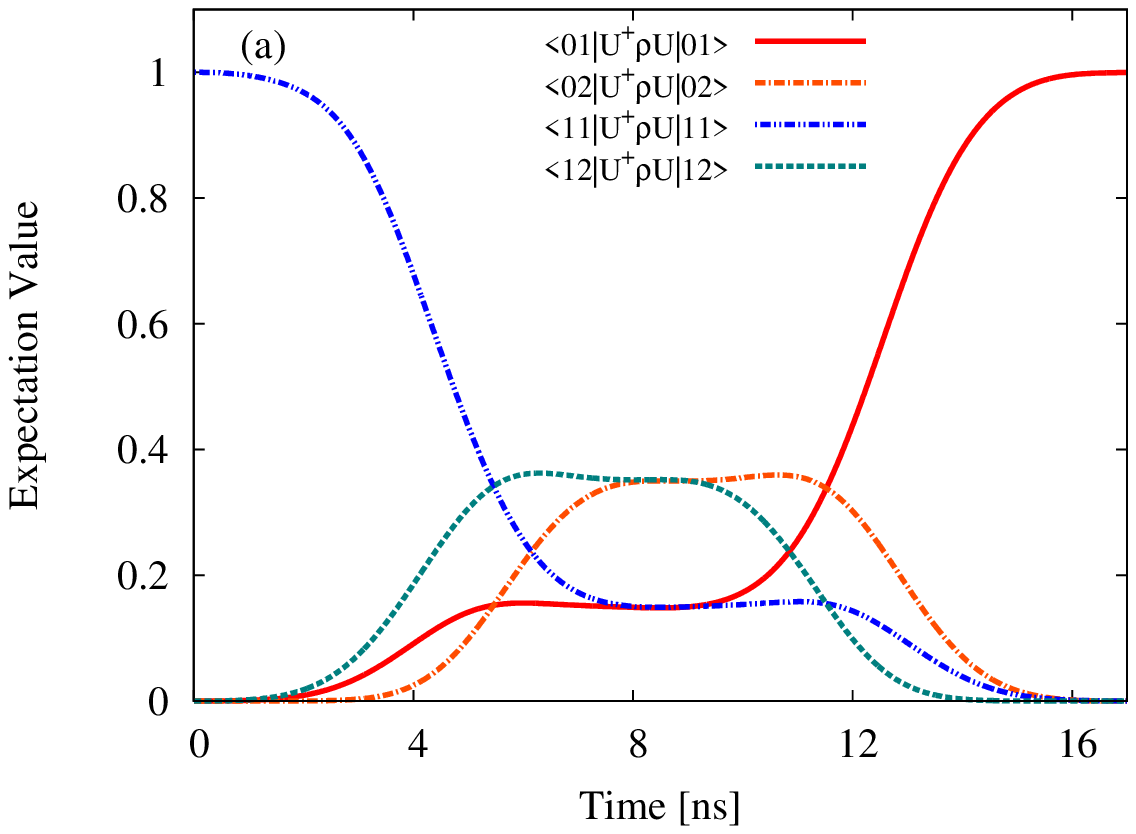}}
  \subfigure{\label{Fig:subfigB}\includegraphics[width=0.95\columnwidth]{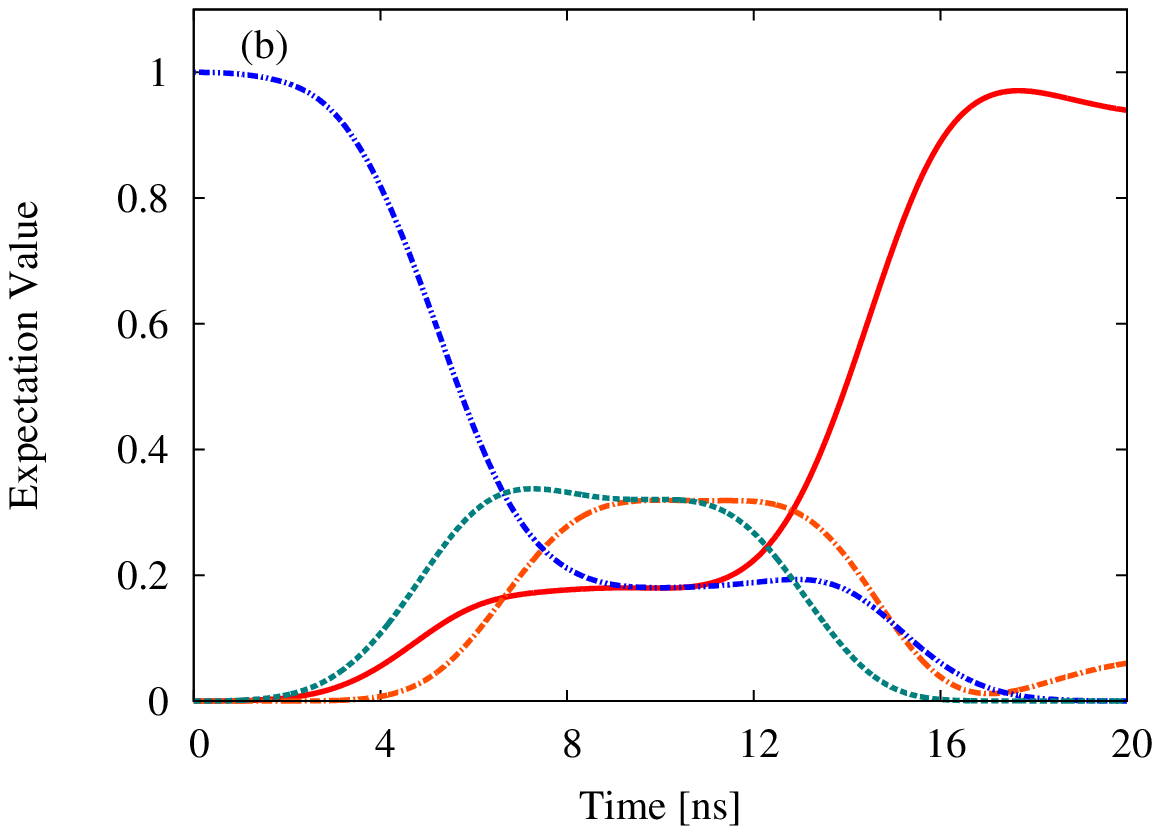}}
  \caption{Populations of the states during the pulse sequence of equation (\ref{eq:wahwah}) for gate time of $17\ns$ \subref{Fig:subfiga}, and $20\ns$ \subref{Fig:subfigB}. At $20\ns$ the  pulse sequence clearly leaves part of the excitation in the $\{\ket{1},\ket{2}\}$ subspace of qubit two, while at $17\ns$ the trajectory is optimal in the sense that no net population transfer is present on qubit two. \label{fig:populs}}
\end{figure}

The state populations during the pulse reveal the underlying mechanism. Figure \ref{fig:populs} shows the populations for gate times $17$ and $20~\ns$. In the latter there is still a net population in the $\ket{2}$ state of the qubit 2 after the gate. For the former, there is no net change to the second qubit at the end. This suggest that the the drive on the second qubit makes it perform a closed transition cycle in the $(\ket{1},\ket{2})$ subspace, thus  acquiring a local phase.

Finally we note in this section that the method worked out here is not the only way to determine new analytical results for pulse shapes. In general, the different terms of equation (\ref{eq:MagnusExpansion}) need to combine into the correct gate in some manner, whereas we have enforced that this combination consists of all terms beyond the lowest one to vanish. Our approach has the advantage that it produces an intuitive result,  providing frequency selectivity criteria eqs. (\ref{eq:FC1},\ref{eq:FC2},\ref{eq:FC3},\ref{eq:FC4}) in the form of the Fourier transform of the driving pulse. 

\subsection{Phase correction}
\label{subsec:phasecor}

\begin{figure}[htbp!] \centering
\includegraphics[width=0.95\columnwidth]{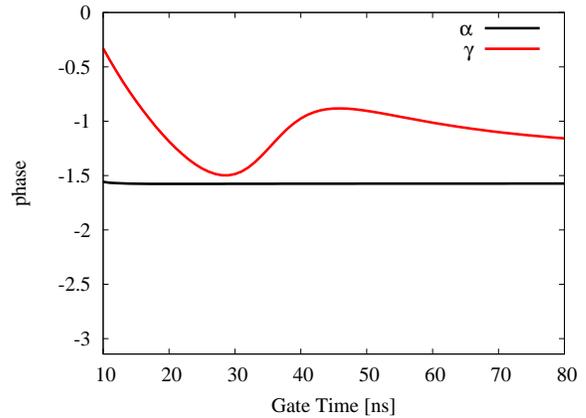}
\caption{Phases as defined in equation (\ref{eq:Udesfidavg}) of the gate with the control sequence from equation (\ref{eq:wahwah}). It is by these phases that the qubits or the subsequent gates need to be corrected.}
\label{fig:wahwahphase}
\end{figure}

The average reduced fidelity (\ref{eq:fidavg}) is insensitive to the phase of the second qubit and leads to a gate of the form of eq. (\ref{eq:Udesfidavg}).
This phase error does not influence population measurements after the gate; only the $X$ and $Y$ component have different contributions. The global phase $\alpha$ and the phase error $\gamma$ for specific gate times are plotted in fig. \ref{fig:wahwahphase}. 
 One can correct for this error in multiple ways. If there is a $Z$ control available on the separate qubits \cite{Paik_PRL_107_240501}  one can simply compensate the phase following
\begin{equation}
\begin{split}
\frac{\pi}{2}=\int Z_1\left(t\right)\mathrm{d}t \\
\alpha\left(t_g\right)=\int Z_2\left(t\right)\mathrm{d}t
\end{split}
\end{equation}
Instead of compensating the qubit phase, one can adjust the phase of the next gate in the $XY$-plane accordingly. This is possible because the phase error is constant given a set gate time, as shown in figure \ref{fig:wahwahphase}. In essence this is the same as changing the frame in the $XY$ plane according to
\begin{equation}
\begin{split}
X'&=\phantom{-}\cos\left(\alpha\left(t_g\right)\right)X+\sin\left(\alpha\left(t_g\right)\right)Y\\
Y'&=-\sin\left(\alpha\left(t_g\right)\right)X+\cos\left(\alpha\left(t_g\right)\right)Y.
\end{split}
\end{equation}
This technique is analogous to phase ramping as described in Refs.\ \cite{Motzoi_PRL_103_110501,Gambetta_PRA_83_012308}
The phases in the leakage states are irrelevant, it is thus sufficient to correct the computational subspaces of the qubits individually. 

\subsection{Experimental protocol \label{sec:implement}}
The procedure to implement the pulse on an actual experiment is
\begin{itemize}
 \item Use spectroscopy to determine the qubit frequencies, yielding $\delta$ and $\Delta$.
 \item Equation (\ref{eq:wahwah}) gives the shape of the pulses for all possible gate times $t_g$. The normalization parameter $A_\pi$ is chosen so that the area theorem, equation (\ref{eq:FC1}), is satisfied, which in general requires numerical root finding.
 \item The gate time $t_g$ is chosen so that the pulse sequence optimizes the reduced average fidelity defined by equation (\ref{eq:fidavg}).
 \item With the gate time known, the phase offset $\alpha(t_g)$ is computed, so that it can be corrected according to the procedures given in section \ref{subsec:phasecor}.
\end{itemize}

\section{Numerical optimized controls}
\label{sec:num}

By using numerical methods one can go beyond the analytic methods discussed in the last sections. Here is discussed how further improvements can be made with the GRAPE algorithm.

\subsection{GRAPE}

To handle our system numerically we use the GRadient Ascent Pulse Engineering (GRAPE) algorithm \cite{Khaneja05}. GRAPE maximizes the fidelity eq. (\ref{eq:fid}) by changing the control amplitudes at discrete times. In discrete time the evolution operator is given by $\hat U\left( t_g\right)=\prod_j \hat U_j$, with $\hat U_j=\mathrm{exp} [-i \hat H\left(j \Delta t \right) \Delta t]$. The fidelity is increased by updating the controls in the direction of the gradient $\Omega_l\left(j\right)=\Omega_l^j \rightarrow \Omega_l^j +\epsilon \partial \Phi / \partial \Omega_l^j$. An analytic expression for the gradient is given in ref. \cite{Machnes11}.

\subsection{Numerical results}
\begin{figure}[htbp!] \centering
\includegraphics[width=0.95\columnwidth]{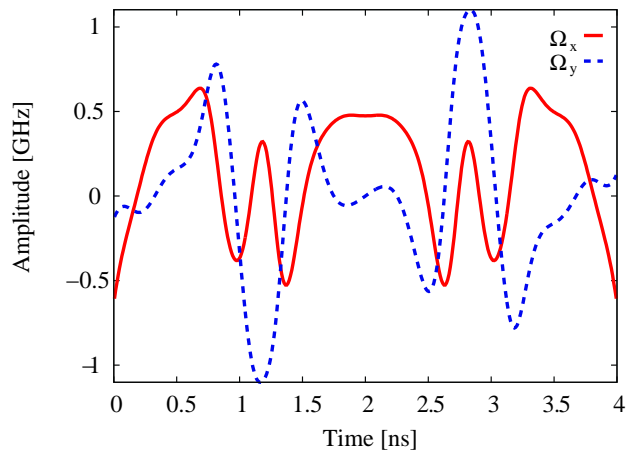}
\caption{Example of a numerically optimized pulse for gate time $t_g=4\ns$ and $\Delta t=0.01\ns$. The pulses for shorter gate time are highly oscillating. The $\Omega_Y$ control is usually not proportional to the derivative of $\Omega_X$.}
\label{fig:controlsfast}
\end{figure}

The system of equation (\ref{eq:HR}) is numerically optimized using the parameters in table \ref{table:2}. Figure \ref{fig:controlsfast} is an example of a short gate  ($4~\mathrm{ns}$) high fidelity ($99.999\%$) GRAPE pulse. This pulse has $t_g\ll\pi/\delta$ and therefore the smallest spectral crowding frequency scale $\delta$ does not impose a quantum speed limit. The limit rather seems to be set by the number of control parameters available. E.g., we have verified that if the size of a time step is 1 ns as in current experimental equipment, the shortest possible time is $8\ns$. From numerical results we have not observed a quantum speed limit. By decreasing the gate time the pulse can be shortened at the expense of higher amplitudes. The pulse in figure \ref{fig:controlsfast} has large amplitudes at $t=0$ and $t=t_g$. These can be removed by adding penalties to the fidelity used by GRAPE \cite{PhysRevB.79.060507}. Only a small increase in gate time is usually needed to enforce pulse sequences to start and end at zero amplitude.
The numerical results show that no speed limit is set by the overlap of the control field in the frequency domain with different qubit transitions. Additionally, numerical pulse sequences don't leave a phase error on the second qubit, eliminating the need for post-processing.

\begin{figure}[htbp!] \centering
\includegraphics[width=0.95\columnwidth]{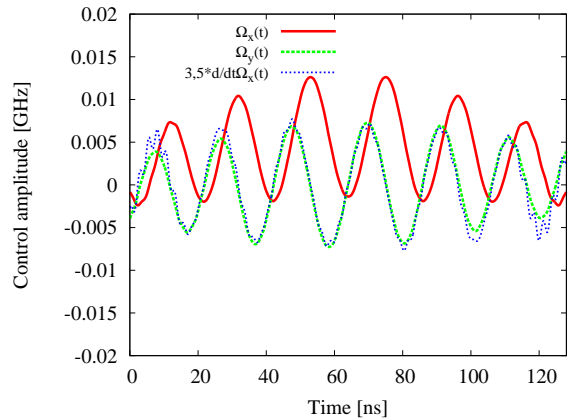}
\caption{Solution found by GRAPE for a long gate time, here $\Delta t=0.01\mathrm{ns}$ and $t_g=130\ns$. The dotted line shows a rescaled version of the derivative of the $\Omega_X$ control. }
\label{fig:controlF}
\end{figure}

To get insight for the shape of the solutions we run the GRAPE algorithm for short time steps and longer gate times to increase the resolution of the discrete time Fourier transform (DTFT). These solutions show rapid oscillations, figure \ref{fig:controlF}. The DTFT of the pulse sequence shows that both quadrature components have contributions at the energy splittings $\delta, \: \delta-\Delta,\: \Delta, 2\delta-\Delta$. This shows that the numerical solution augments the 
one based on the Magnus expansion by adding small further sideband drives.

When one goes to shorter gate times however Fourier analysis shows that the contribution of the higher frequency components increases, making the Fourier transform less useful due to the lower frequency resolution. For faster pulses one could suggest that adding more sideband modulations could improve the results further.

\begin{figure}[htbp!] \centering
\includegraphics[width=0.95\columnwidth]{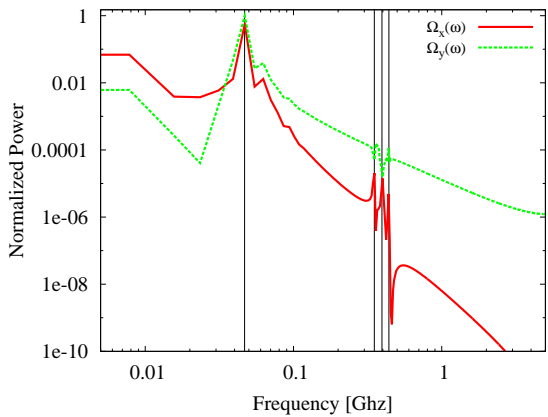}
\caption{Fourier transform of the pulse shown in \ref{fig:controlF} found by GRAPE. }
\label{fig:controlFT}
\end{figure}

\section{Conclusion}

We have found numerical as well as analytical pulse shapes implementing single qubit gates in a 3D cavity coupled to two single junction Transmons. Such qubits are typically hindered by spectral crowding whereby leakage transitions lie close in frequency to main qubit $0\leftrightarrow1$ transitions. 
We combine average Hamiltonian theory for arbitrary waveforms with the DRAG methodology, shows that it is possible to find better controls using a sideband modulation. 

Numerically optimized pulses support this conclusion and provide greater improvements in fidelity. They show that qubits can still be addressed individually with short gate times. Faster control pulses require more bandwidth and amplitude, therefore the limiting factor is the capabilities of the arbitrary waveform generator. No speed limit has been observed in numerically optimized pulses, which is contrary to the believe that spectral crowding limits the scalability of the 3D cavity architecture in cQED.

We would like to thank Leo DiCarlo for suggesting this problem as well as Felix Motzoi for useful discussions. This work was supported through IARPA within the MQCO program and the European Union within SCALEQIT.

\bibliographystyle{apsrev}
\bibliography{TransmonXGate}{}

\end{document}